\documentstyle[epsf]{mn}

\newif\ifAMStwofonts

\def\logz{\lbrack\hbox{Fe/H}\rbrack}

\def\eps@scaling{.95}
\def\epsscale#1{\gdef\eps@scaling{#1}}
\def\plotone#1{\centering \leavevmode
    \epsfxsize=\eps@scaling\columnwidth \epsfbox{#1}}

\ifoldfss
  \ifCUPmtlplainloaded \else
    \NewTextAlphabet{textbfit} {cmbxti10} {}
    \NewTextAlphabet{textbfss} {cmssbx10} {}
    \NewMathAlphabet{mathbfit} {cmbxti10} {} 
    \NewMathAlphabet{mathbfss} {cmssbx10} {} 
  \fi
  \ifAMStwofonts
    \ifCUPmtlplainloaded \else
      \NewSymbolFont{upmath} {eurm10}
      \NewSymbolFont{AMSa} {msam10}
      \NewMathSymbol{\upi}     {0}{upmath}{19}
      \NewMathSymbol{\umu}     {0}{upmath}{16}
      \NewMathSymbol{\upartial}{0}{upmath}{40}
      \NewMathSymbol{\leqslant}{3}{AMSa}{36}
      \NewMathSymbol{\geqslant}{3}{AMSa}{3E}

    \fi
  \fi
\fi 

\ifnfssone
  \newmathalphabet{\mathit}
  \addtoversion{normal}{\mathit}{cmr}{m}{it}
  \addtoversion{bold}{\mathit}{cmr}{bx}{it}
  \newmathalphabet{\mathbfit} 
  \addtoversion{normal}{\mathbfit}{cmr}{bx}{it}
  \addtoversion{bold}{\mathbfit}{cmr}{bx}{it}
  \newmathalphabet{\mathbfss} 
  \addtoversion{normal}{\mathbfss}{cmss}{bx}{n}
  \addtoversion{bold}{\mathbfss}{cmss}{bx}{n}
  \ifAMStwofonts
    \ifCUPmtlplainloaded \else
      \UseAMStwoboldmath
      \makeatletter
      \new@mathgroup\upmath@group
      \define@mathgroup\mv@normal\upmath@group{eur}{m}{n}
      \define@mathgroup\mv@bold\upmath@group{eur}{b}{n}
      \edef\UPM{\hexnumber\upmath@group}
      \new@mathgroup\amsa@group
      \define@mathgroup\mv@normal\amsa@group{msa}{m}{n}
      \define@mathgroup\mv@bold\amsa@group{msa}{m}{n}
      \edef\AMSa{\hexnumber\amsa@group}
      \makeatother
      \mathchardef\upi="0\UPM19
      \mathchardef\umu="0\UPM16
      \mathchardef\upartial="0\UPM40
      \mathchardef\leqslant="3\AMSa36
      \mathchardef\geqslant="3\AMSa3E
    \fi
  \fi
\fi 

\ifnfsstwo
  \DeclareMathAlphabet{\mathbfit}{OT1}{cmr}{bx}{it}
  \SetMathAlphabet\mathbfit{bold}{OT1}{cmr}{bx}{it}
  \DeclareMathAlphabet{\mathbfss}{OT1}{cmss}{bx}{n}
  \SetMathAlphabet\mathbfss{bold}{OT1}{cmss}{bx}{n}
  \ifAMStwofonts
    \ifCUPmtlplainloaded \else
      \DeclareSymbolFont{UPM}{U}{eur}{m}{n}
      \SetSymbolFont{UPM}{bold}{U}{eur}{b}{n}
      \DeclareSymbolFont{AMSa}{U}{msa}{m}{n}
      \DeclareMathSymbol{\upi}{0}{UPM}{"19}
      \DeclareMathSymbol{\umu}{0}{UPM}{"16}
      \DeclareMathSymbol{\upartial}{0}{UPM}{"40}
      \DeclareMathSymbol{\leqslant}{3}{AMSa}{"36}
      \DeclareMathSymbol{\geqslant}{3}{AMSa}{"3E}
    \fi
  \fi
\fi 

\ifCUPmtlplainloaded \else
  \ifAMStwofonts \else 
    \def\upi{\pi}
    \def\umu{\mu}
    \def\upartial{\partial}
  \fi
\fi

\title[The Stellar Content and Distance of UGC 4483]{The Stellar Content and Distance of UGC 4483\thanks{Based on observations with the NASA/ESA \textit{Hubble Space Telescope}, obtained at the Space Telescope Science Institute, which is operated by the Association of Universities for Research in Astronomy, Inc., under NASA contract NAS 5--26555. These observations are associated with proposal ID GO--8192.}}
\author[A. E. Dolphin et al.]
 {A. E. Dolphin,$^1$
 L. Makarova,$^2$
 I. D. Karachentsev,$^2$
 V. E. Karachentseva,$^3$
\newauthor D. Geisler,$^4$
 E. K. Grebel,$^5$
 P. Guhathakurta,${^6}{^7}$
 P. W. Hodge,$^8$
 A. Sarajedini,$^9$
\newauthor P. Seitzer$^{10}$ \\
$^1$Kitt Peak National Observatory, National Optical Astronomy Observatories, PO Box 26732, Tucson, AZ, 85726, USA \\
electronic mail: dolphin@noao.edu \\
$^2$Special Astrophysical Observatory, Russian Academy of Sciences, N. Arkhyz, KChR, 357147, Russia \\
$^3$Astronomical Observatory of Kiev University, 04053, Observatorna 3, Kiev, Ukraine \\
$^4$Departmento de Fisica, Grupo de Astronomia, Universidad de Concepcion, Casilla 160-C, Concepcion, Chile \\
$^5$Max-Planck-Institut f\"ur Astronomie, K\"onigstuhl 17, D-69117, Heidelberg, Germany \\
$^6$UCO/Lick Observatory, University of California at Santa Cruz, Santa Cruz, CA, 95064, USA \\
$^7$Alfred P. Sloan Research Fellow \\
$^8$Department of Astronomy, University of Washington, P.O. Box 351580, Seattle, WA, 98195, USA \\
$^9$Astronomy Department, Wesleyan University, Middletown, CT, 06459, USA \\
$^{10}$Department of Astronomy, University of Michigan, 830 Dennison Building, Ann Arbor, MI, 48109, USA}

\date{Accepted . Received }

\pagerange{\pageref{firstpage}--\pageref{lastpage}}
\pubyear{1994}

\begin{document}

\maketitle

\label{firstpage}

\begin{abstract}
We present HST/WFPC2 observations of UGC 4483, an irregular galaxy in the M81/NGC 2403 complex. Stellar photometry was carried out with HSTphot, and is complete to $V \simeq 26.0$ and $I \simeq 24.7$. We measure the red giant branch tip at $I = 23.56 \pm 0.10$ and calculated a distance modulus of $\mu_0 = 27.53 \pm 0.12$ (corresponding to a distance of $3.2 \pm 0.2$ Mpc), placing UGC 4483 within the NGC 2403 subgroup. We were able to measure properties of a previously-known young star cluster in UGC 4483, finding integrated magnitudes of $V = 18.66 \pm 0.21$ and $I = 18.54 \pm 0.10$ for the stellar contribution (integrated light minus H$\alpha$ and $\lbrack$OIII$\rbrack$ contribution), corresponding to an age of $\sim 10-15$ Myr and initial mass of $\sim 10^4$ $M_{\odot}$. This is consistent with the properties of the cluster's brightest stars, which were resolved in the data for the first time. Finally, a numerical analysis of the galaxy's stellar content yields a roughly constant star formation rate of $1.3 \times 10^{-3} M_{\odot}$yr$^{-1}$ and mean metallicity of $\logz = -1.3$ dex from 15 Gyr ago to the present.
\end{abstract}


\section{INTRODUCTION}
A dwarf irregular galaxy, UGC 4483 is situated midway on the sky between the bright spirals M81 and NGC 2403 ($7^{\circ}$ from both). Its location and heliocentric radial velocity (+156 km s$^{-1}$, Thuan \& Seitzer 1979) indicate that it is a potential member of the M81 / NGC 2403 complex of galaxies. This was confirmed through photometry of its brightest blue stars by Tikhonov \& Karachentsev (1992), who found a distance of $3.6 \pm 0.7$ Mpc ($\mu_0 = 27.8 \pm 0.4$ magnitudes).
\begin{figure*}
\epsscale{2.0}
\plotone{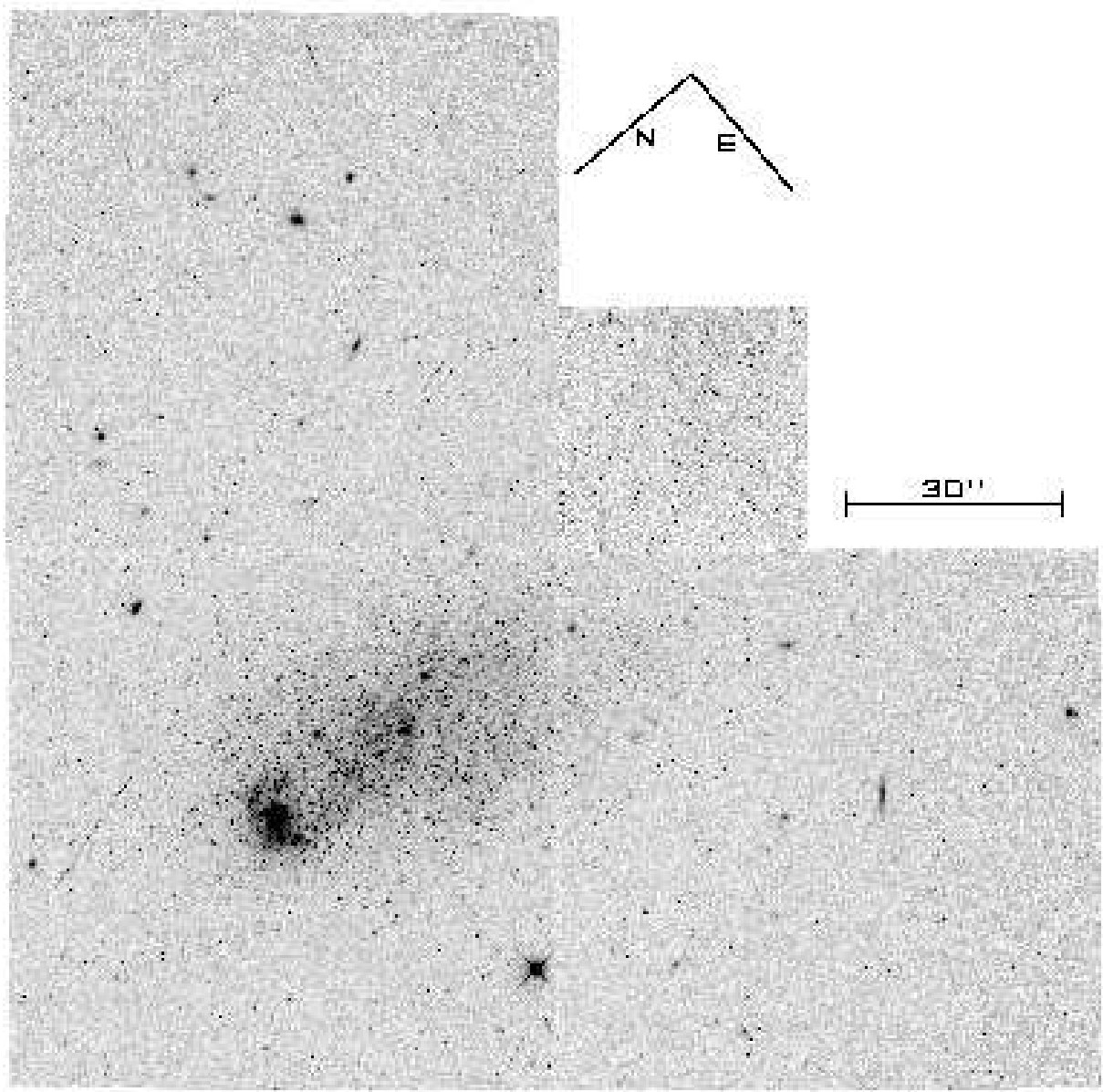}
\epsscale{0.95}
\caption{Cosmic ray-cleaned F606W image of UGC 4483. North and East are as shown; the length of the bar in the upper-right is 30 arcsec.}
\label{figImage}
\end{figure*}

Integrated magnitudes of UGC 4483 were measured by Bremnes et al. (1998) in their CCD observations of potential M81 group members, and determined to be $B = 15.12$ and $R = 14.70$. A similar study by Makarova (1999) produced integrated magnitudes of $B = 14.95$, $V = 14.51$, and $I = 14.02$. Given the difficulty of calculating integrated light and the presence of a bright star 3 arc-minutes to the south, these values are in agreement.

Karachentseva, Karachentsev, \& Boerngen (1985) presented large-scale photographs of dwarf galaxies in the M81 region, and found a bright knot in the northern part of UGC 4483. Spectra by Skillman (1991) and Izotov et al. (1992) showed this to be an HII region, with similar metallicity to I Zw 18 ($\logz \simeq -1.4$).

In this paper, we present the first \textit{Hubble Space Telescope} (\textit{HST}) observations of UGC 4483. The superior resolution of these data allows us to study the stellar content of the galaxy, enabling a more robust distance via the RGB tip as well as a first study of the stellar content.

Section 2 describes our observations and photometry. Section 3 focuses on the galaxy's integrated properties, while section 4 concentrates on the stellar content.

\section{OBSERVATIONS AND REDUCTIONS}

\subsection{Observations}
Two WFPC2 images of UGC 4483 were obtained on 24 January 2000 as part of program GO--8192, the dwarf galaxy snapshot survey (Seitzer et al. 1999). The pointing was made so that the galaxy center (RA = 8$^h$37$^m$03.0$^s$, Dec = +69$^{\circ}$46'31'', J2000) would be centered in WFC3. As with all other objects imaged by the program, the exposures were 600 seconds each, taken in F606W and F814W. Images were obtained from the STScI archive using the standard processing and calibration pipeline.

\subsection{Reduction}
Reduction was made using the HSTphot stellar photometry package (Dolphin 2000a). Before running photometry, the data quality images were used to mask bad and possibly-bad pixels, and pixels containing bright cosmic rays were masked using the HSTphot \textit{cleansep} utility. \textit{Cleansep} is a program that cleans cosmic rays by comparing the two images (F606W and F814W) similarly to the IRAF task CRREJ, except that it must allow for a range of possible $(V-I)$ colours because the two images are taken with different filters. Rejected pixels were masked out from the remainder of the reductions. (Normally, one would prefer to clean using a comparison of two images with identical exposure time and filter, but that was not possible in this case, given the requirements of a snapshot proposal.) The F606W mosaic image, cleaned in this way, is shown in Figure \ref{figImage}.

The stellar photometry was carried out with \textit{multiphot}, a multiple image processing program similar in concept to ALLFRAME (Stetson 1994). For each chip, the images were aligned to an accuracy of 0.1--0.2 PC pixels, and the F606W and F814W photometry made simultaneously. Because of the crowding in the main body of the galaxy, we used the \textit{multiphot} option to recalculate the sky level within a small annulus surrounding the star during each photometry iteration.

Although HSTphot uses a library of model PSFs based on Tiny Tim, changes in the focus of \textit{HST} require that the model PSFs be corrected. Normally this is done using the residuals of bright, isolated stars, but the lack of such stars in WFC2 made such a task impossible for these data. Instead, a semi-empirical correction was made by requiring that the median \textit{sharpness} value of the stars in each chip of each image be zero. Examination of the residuals shows that, while not perfect, the systematic errors in faint star photometry caused by PSF errors should be no more than 0.015 magnitudes, significantly less than the standard errors in their photometry.

The final photometry was then made using the corrected PSF library, and aperture corrections and the Dolphin (2000b) charge-transfer inefficiency correction and calibration applied. We estimate the aperture corrections in the three wide field chips to be accurate to 0.05 magnitudes in WFC2, 0.02 in WFC3, and 0.03 in WFC4. Because of the small sky coverage of the planetary camera (PC), lack of stars for an accurate aperture correction, and chip's distance from the centre of UGC 4483, the PC photometry is omitted from our results. Additionally, stars with $\hbox{signal-to-noise} < 5$, $\chi > 2.0$, or $|\hbox{sharpness}| > 0.4$ in either exposure were eliminated from the final photometry list, in order to minimize the number of false detections. Artificial star tests were also made, so that the accuracy and completeness of the (crowded) WFC3 photometry could be measured.

\begin{figure}
\plotone{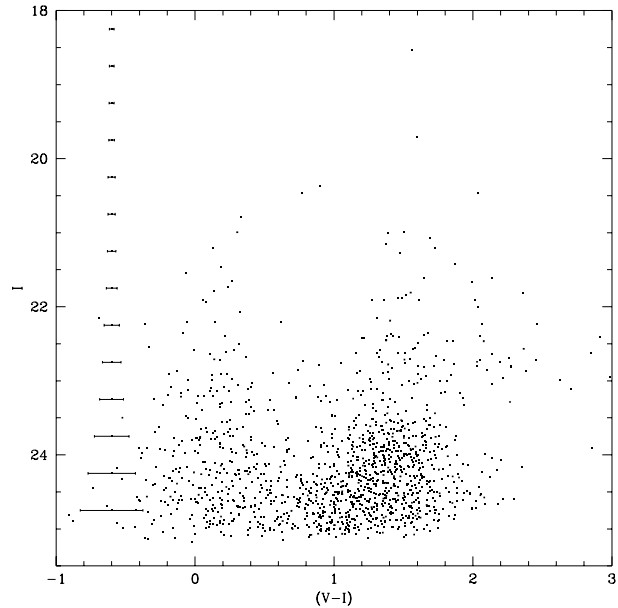}
\caption{CMD of 1378 stars in UGC 4483, using all three WFC chips. Mean uncertainties in $(V-I)$ as a function of $I$ magnitude are shown along the left-hand side.}
\label{figCMD}
\end{figure}
\begin{figure}
\plotone{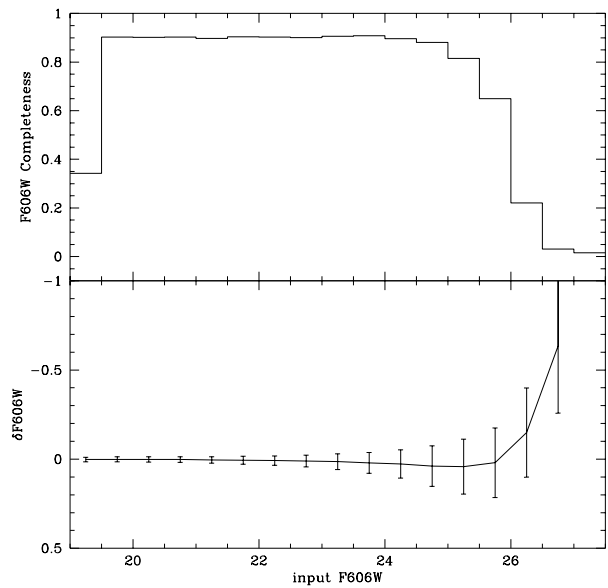}
\caption{F606W completeness and photometric errors. In the bottom panel, the line shows the mean output minus input magnitudes of artificial stars, and the error bars show the $1 \sigma$ distribution}
\label{figVcmplt}
\end{figure}
\begin{figure}
\plotone{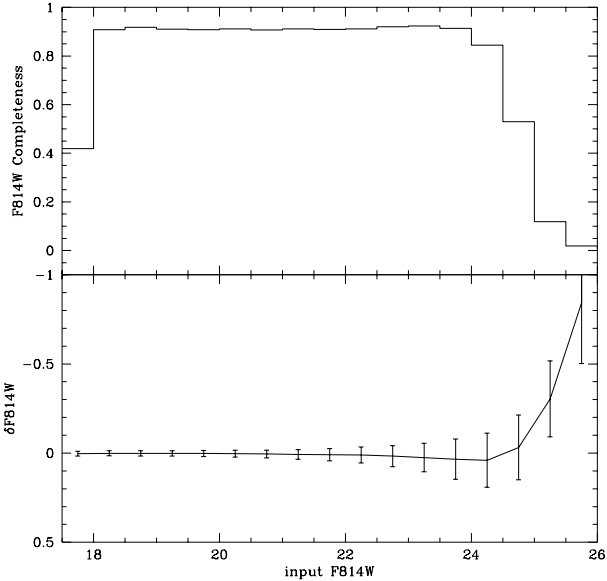}
\caption{Like Figure \ref{figVcmplt}, but for F814W}
\label{figIcmplt}
\end{figure}
Figure \ref{figCMD} shows the colour-magnitude diagram (CMD) of the WFC data; completeness and mean photometric error (due to crowding and faint-end bias) in WFC3 are shown in Figures \ref{figVcmplt} and \ref{figIcmplt}, respectively. The $\sim$90\% maximum completeness level in these reductions is primarily a result of the fraction of stars falling within one pixel of a masked pixel (cosmic ray, bad pixel, bad column, etc) and thus rejected. This is a reasonable value, as the WFPC2 exposure time calculator indicates that roughly 4\% of the pixels should be affected by cosmic rays for these observations.

\section{DISTANCE AND GROUP MEMBERSHIP}

\subsection{Distance}
As demonstrated by Lee, Freedman, \& Madore (1993), the red giant branch (RGB) tip is a useful distance indicator that is relatively independent of age and metallicity. From our CMD, we find the RGB tip to fall at $I = 23.56 \pm 0.10$ and $(V-I) = 1.45 \pm 0.14$, using both a Sobel filter and a first derivative test. After correcting for extinction using the maps of Schlegel, Finkbeiner, \& Davis (1998), these values become $I_0 = 23.49 \pm 0.10$ and $(V-I)_0 = 1.40 \pm 0.14$. From Skillman's (1991) spectroscopy, we estimate a metallicity of $\logz = -1.4 \pm 0.2$.

We determined the absolute magnitudes using two methods, the semi-empirical calibration given by Lee et al. (1993) and the theoretical isochrones of Girardi et al. (2000). The semi-empirical calibration, adopting the above values for dereddened colour and metallicity, results in an absolute magnitude of $M_I = -4.08 \pm 0.05$. Using the RGB tip absolute magnitudes for a range of ages from the Girardi et al. (2000) isochrones for Z=0.001 and Z=0.0004, we find an absolute magnitude of $M_I = -4.00 \pm 0.04$. Taking the mean of the two values and adding half the difference to the uncertainty, we adopt an absolute magnitude of $M_I = -4.04 \pm 0.06$ for the UGC 4483 RGB tip. This value results in a true distance modulus of $\mu_0 = 27.53 \pm 0.12$, corresponding to a distance of $3.2 \pm 0.2$ Mpc.

\subsection{Membership}
Karachentsev et al. (2000) find distance moduli of $\mu_0 = 27.84 \pm 0.05$ for eleven galaxies in the M81 group and $\mu_0 = 27.50 \pm 0.02$ for two galaxies in the NGC 2403 subgroup. Our distance modulus of $\mu_0 = 27.53 \pm 0.12$ for UGC 4483 places it firmly in the NGC 2403 subgroup, with a distance of $0.39 \pm 0.08$ Mpc from NGC 2403. The case for NGC 2403 subgroup membership is strengthened by radial velocity data, with the RC3 catalog (de Vaucouleurs et al. 1991) listing heliocentric velocities of $-34 \pm 4$ km s$^{-1}$ for M81, $+131 \pm 3$ km s$^{-1}$ for NGC 2403, and $+156 \pm 5$ km s$^{-1}$ for UGC 4483.

\section{STELLAR CONTENT}

\subsection{Populations}
\begin{figure}
\plotone{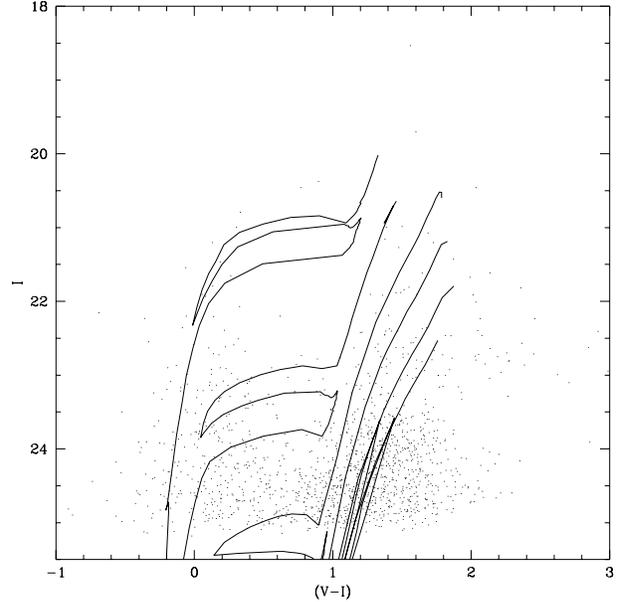}
\caption{CMD of UGC 4483, with isochrones of $\logz = -1.4$ overplotted. Isochrones are for log(age) = 7.5, 8.0, 8.5, 9.0, 9.5, and 10.0, and are interpolated from Girardi et al. (2000).}
\label{figISOCMD}
\end{figure}
Figure \ref{figISOCMD} shows the CMD of UGC 4483, with isochrones (again with the isochrones interpolated from the Girardi et al. 2000 isochrones to metallicity $\logz = -1.4$ based on Skillman 1991) overplotted. Although a more detailed star formation history will be determined below via a numerical approach, we find it helpful to use this diagram to analyse the stellar populations present. As a first ``sanity check'' on our photometry, distance, and metallicity estimate, we see that the isochrones overlap the observed CMD quite well, with the blue loops (core helium-burning stars), red supergiants, and RGB all falling where they are expected. The photometry does suggest some metallicity spread, especially in the metal-rich direction, as these isochrones (even if photometric error is added) do not account for the width of the RGB tip. We also note the presence of comparatively metal-rich AGB stars (above and red of the RGB tip), very similar to those found in UGC 6456 (VII Zw 403), a similar galaxy, by Lynds et al. (1998). (It should be noted that, while our data are not particularly deep, this feature falls well above the photometric cutoff, and is thus not a result of noise.)

In our qualitative star formation history, we first turn our attention to the young populations along the blue loops and upper main sequence.  At the metallicity of UGC 4483, we observe that it is difficult to unambiguously divide these two populations. However, the upper main sequence should be generally blueward of $(V-I) = 0$ and the blue loop stars redward of this value, and the CMD does show a hint of a division along that line. If this break is real, than the brightest main sequence star is at $I = 22.2$, corresponding to an absolute magnitude of $M_I = -5.4$ and an age of $\sim 10$ Myr.  This age of most recent star formation is also consistent with the brightest blue supergiants, which require an age of most recent star formation of $\sim 15$ Myr. The isochrones also demonstrate that the blue loop sequence amounts to an age sequence, and we can trace the persistence of star formation from the recent age of $\sim 10$ Myr to the photometric cutoff near $\sim 200$ Myr. From $\sim 200$ to $\sim 300$ Myr, the position of the blue loop stars falls in the photometric noise, giving evidence that such stars are present but no detailed information.

Between $\sim$300 Myr (the last trace of blue loop stars in our observed CMD) and $\sim$2 Gyr (the onset of the RGB), we have very little trace of the star formation in the CMD; only the rapid red supergiant phase. However, we do see a statistically significant number of stars in that region, implying that there was star formation during this period.

Finally, there is a significant old ($> 2$ Gyr) population, as evidenced by the RGB. As these data lack age-sensitive features such as the red clump and horizontal branch, we would be hard-pressed to identify specific star forming ages during that period. However, we do see relatively faint, metal-rich ($\logz \simeq -1.0$) AGB stars, which would imply that there was star formation roughly 2--5 Gyr ago. The evidence of the presence of ``metal-rich'' stars is also strengthened by the red extent of the RGB tip, whose dereddened colour of $(V-I)_0 = 1.7$ corresponds to a metallicity of $\logz \simeq -0.9$.

In short, the main sequence and blue loop phases show evidence of continuous star formation from roughly 10 to 200 Myr ago at low metallicity ($\logz \simeq -1.4$). The RGB and AGB give evidence of star formation over a metallicity range (as rich as $\logz \simeq -1.0$) for stars older than 2 Gyr ago. From the AGB, there is conclusive evidence of star formation roughly 3 Gyr ago; otherwise the only statement we can make with certainty is that UGC 4483 had star formation prior to 2 Gyr ago, based on the presence of an RGB. The red supergiants help fill in the gap between the faintest blue loop stars and the youngest RGB stars, again providing evidence for some star formation during that period but no detailed information.

\begin{figure}
\plotone{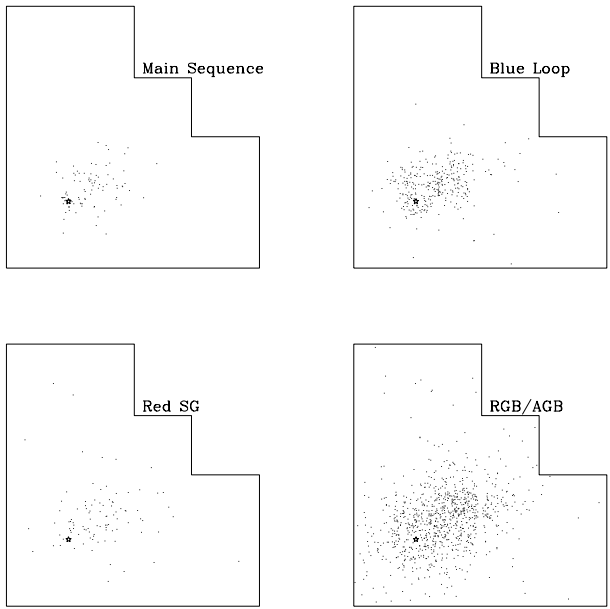}
\caption{Spatial distribution of the stellar populations in UGC 4483. The main sequence region is bounded by $(V-I) < 0$ and $I > 22$; all other stars bluer than $(V-I) = 0.75$ are classified as blue loop stars. The red supergiant region is bounded by $(V-I) > 0.75$, $I < 23$, and blue of the 2 Gyr isochrone; all other stars red of $(V-I) = 0.75$ are classified as RGB or AGB stars. For reference, the position of the cluster is indicated by the star in each panel.}
\label{figDist}
\end{figure}
The spatial distribution of the various populations is shown in Figure \ref{figDist}. The four plots correspond to an age sequence -- main sequence (roughly 10--50 Myr), blue loop (15--300 Myr), red supergiant (0.1--2 Gyr), and RGB/AGB ($>2$ Gyr). The oldest stars show very little structure, with the density of points smoothly tracing the galaxy's overall structure (the ``hole'' at lower left of the galaxy body is due to the incompleteness caused by the presence of a massive star cluster). The blue loop and main sequence stars, however, are concentrated in the main body of the galaxy, with the main sequence stars concentrated heavily near the young star cluster.

\subsection{Star Cluster}
As observed in the Karachentseva et al. (1985) photographic images, UGC 4483 has a star cluster on its northern edge, which shows clearly in Figure \ref{figImage} in the lower-left part of the galaxy. The resolution of the present data permit us to study the cluster with minimal contamination from the galaxy. From the F606W image, we estimate the cluster to be centered at approximately X=425.5, Y=417.5 on WFC3 (RA = 8$^h$37$^m$02.8$^s$, Dec = +69$^{\circ}$46'51'', J2000), with a radius of roughly 30 pixels (3 arc-sec). At the distance calculated above, this radius corresponds to a linear diameter of roughly 90 pc, significantly larger than a typical young LMC cluster ($< 10$ pc) and smaller than a giant HII region such as 30 Doradus ($\sim$1 kpc).

Integrated photometry was performed on the distortion-corrected and cosmic ray-cleaned frames through aperture photometry. Aperture photometry, using background values calculated from the region surrounding the 30-pixel aperture, produces magnitudes of $\hbox{F606W} = 18.06 \pm 0.13$ and $\hbox{F814W} = 18.55 \pm 0.10$. The relatively large uncertainties in these magnitudes are dominated by the highly variable background. We have chosen to leave these values untransformed because a significant fraction of the F606W counts come from H$\alpha$ and [OIII], and thus the transformations (which assume stellar spectra) will be invalid. Furthermore, while H$\alpha$ falls well within the F606W bandpass, it does not fall within the standard $V$ bandpass, creating additional error in any attempted transformation.

We can attempt to correct for the line contamination, however, using the H$\alpha$ flux and line ratios of Skillman et al. (1994) with the WFPC2 exposure time calculator. We estimate a total line contamination (primarily from H$\alpha$, $\lbrack$OIII$\rbrack$, and H$\beta$) of $33 \pm 7$ DN/sec through F606W. A correction for this flux increases our F606W magnitude to $\hbox{F606W} = 18.62 \pm 0.21$. This value, which should contain only stellar flux, can be transformed with the F814W magnitude above ($\hbox{F814W} = 18.55 \pm 0.10$), producing standard magnitudes of $V = 18.66 \pm 0.21$ and $I = 18.54 \pm 0.10$. Adopting our distance and extinction values from above, we find extinction-corrected absolute magnitudes of $M_V = -8.98 \pm 0.24$ and $M_I = -9.06 \pm 0.16$, corresponding to an age of $\sim 15$ Myr and an initial mass of $\sim 10^4$ $M_{\odot}$ (again using Girardi et al. 2000 isochrones, interpolated to $\logz = -1.4$). For comparison, this is slightly fainter than the cluster inside 30 Doradus, which has an absolute magnitude of $M_V \simeq -9.3$, adopting $V = 9.5$ from SIMBAD and an extinction of $A_V \simeq 0.3$.

\begin{figure}
\plotone{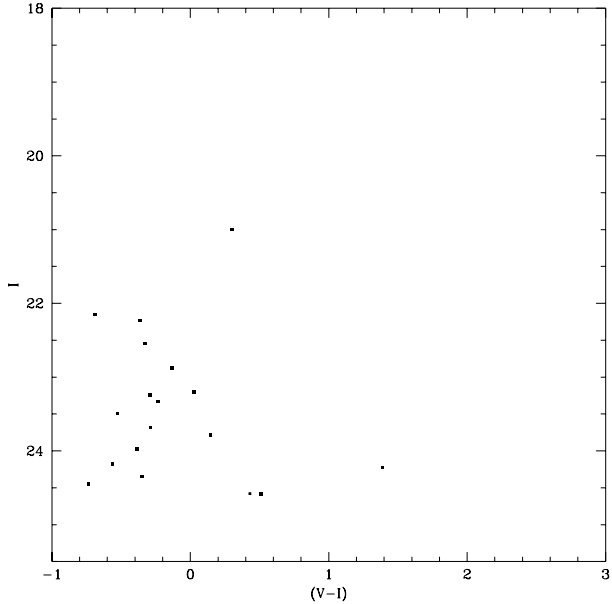}
\caption{CMD of the UGC 4483 cluster.}
\label{figClusCMD}
\end{figure}
Finally, we have photometry for 18 stars falling in this region (within 30 pixels of 425.5, 417.5 on WFC3), and the CMD is shown in Figure \ref{figClusCMD}. Although the quality of the photometry poor because of the crowding, it is clear that this is a young object.  The brightest star is an evolved star (blue supergiant), and corresponds to the second brightest blue star seen in the full CMD shown in Figure \ref{figISOCMD}.  All other stars with good photometry are main sequence stars, with the colour and turnoff corresponding to an age of $\sim10$ Myr.

\subsection{Star Formation History}
Although our photometry of the stars in UGC 4483 contains only the brightest features, we can attempt to make a quantitative measurement of the star forming history of the galaxy, using the population synthesis technique described by Dolphin (2000c). To build our synthetic CMDs, we adopted the isochrones of Girardi et al. (2000) for stars with mass less than $7 M_{\odot}$, previous Padova models for more massive stars, and newly-computed models for $\logz = -2.3$, all kindly provided by Leo Girardi.

\begin{figure}
\plotone{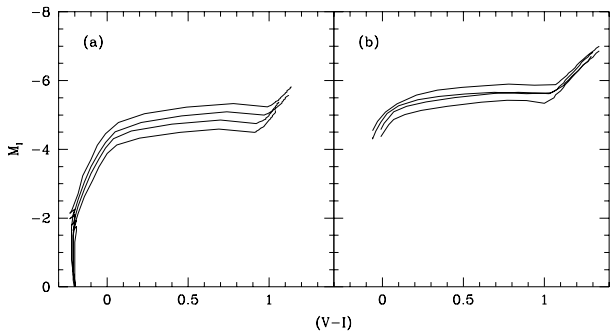}
\caption{Isochrones from log(age) = 7.70 to 7.85, by steps of 0.05 dex. The younger two (log(age) = 7.70 and 7.75, brighter on the diagram) are calculated by interpolation between $\logz$ of $-0.7$ and $-1.7$; the older two (log(age) = 7.80 and 7.85, fainter on the diagram) are taken from Girardi et al. 2000. The left panel shows early phases of evolution (main sequence through the earliest stages of core helium burning); the right panel shows the blue and red supergiant phases.}
\label{figIsoInterp}
\end{figure}
A possible source of significant uncertainty is a ``hole'' in the complete set of isochrones, as no isochrones are available for young stars (log(age) $<$ 7.8) with $\logz$ of $-1.3$ (which happens to be the set of models that would be predominantly used for UGC 4483's young populations). Instead, we have calculated isochrones with $\logz = -1.3$ by interpolating from the adjacent metallicities ($\logz$ of $-0.7$ and $-1.7$). Figure \ref{figIsoInterp} shows isochrones immediately before the break (calculated by interpolation) and after the break (not interpolated), and demonstrates that the transition between the two sets of models is extremely smooth. Thus we conclude that, despite the possible errors from interpolation and discontinuity at log(age) = 7.8, the models used are satisfactory.

Using these isochrones (which we have interpolated for arbitrary metallicities and ages) and a comprehensive library of artificial stars, we have attempted to use a maximum likelihood routine to determine the past star formation rate as a function of time and metallicity. This was done by converting the colour-magnitude to a Hess diagram (with resolution 0.3 magnitudes in $V$ and 0.1 magnitudes in $V-I$), and computing simulated Hess diagrams for various combinations of age and metallicity. Based on the ages corresponding to different populations in our CMD, we chose 4 age ranges for the simulated CMDs: 0--0.1 Gyr (corresponding to the observed main sequence), 0.1--0.3 Gyr (blue loops), 0.3--2.0 Gyr (few observed stars; some red supergiants), and 2.0--15.0 Gyr (RGB and AGB).  We allowed for a range of metallicity from $\logz = -0.7$ to $\logz = -2.3$, with adjacent sets of models spaced by 0.4 dex. Thus for a given distance and extinction, we calculated a total of 25 simulated CMDs and used a maximum likelihood estimator to determine the combination of simulated CMDs that best resembles the observed data. We stress that, while star formation history studies are generally done with much deeper data, we are only attempting to resolve epochs of star formation that correspond to distinct parts of the CMD.

Because the statistics in this problem are Poisson rather than normal, the probability of finding $N$ stars in a CMD bin with $M$ predicted (model) points is
\begin{equation}
P = \frac{M^N}{N! e^M},
\end{equation}
where $M$ is given by the sum of the star formation rate times the number of points for the particular model in the CMD bin,
\begin{equation}
M = \Sigma_t \Sigma_Z SFR(t,Z) M(t,Z).
\end{equation}
Thus the total probability of observing a CMD with $N_i$ observed points per bin and $M_i$ model points per bin is
\begin{equation}
P = \Pi_i \frac{M_i^{N_i}}{N_i! e^{M_i}},
\end{equation}
or
\begin{equation}
\ln P = \Sigma_i N_i \ln M_i - \Sigma_i \ln (N_i!) - \Sigma_i M_i,
\end{equation}
which we wish to maximize. Because changing our trial star formation history alters the number of model points rather than the number of observed points and because we are interested in calculating the model with the maximum probability rather than the probability itself, the $\Sigma_i \ln(N_i!)$ term is independent of the model choice and is omitted for the sake of simplifying the calculation. Additionally, as our numerical minimization routine is written to minimize (rather than maximize) the function in question, we chose to minimize the function
\begin{equation}
f = \Sigma_i (M_i - N_i \ln M_i).
\end{equation}
This function is the Poisson equivalent of the standard $\chi^2$ minimization function
\begin{equation}
\chi^2 = \Sigma_i \frac{(M_i - N_i)^2}{M_i}.
\end{equation}
It is clear that for points with bad data (where $N_i$ is large but no model points are predicted), the fitting function becomes infinite (as would a $\chi^2$ minimization). To compensate, we have added a small number of bad points to every CMD bin, corresponding to 5\% of the points at bright magnitudes and 25\% of the points at the faintest magnitudes. This allows us to account for the fact that the observed CMD contains objects that cannot be accounted for in our model of isochrones plus scatter from artificial star tests -- most significantly, low-level cosmic rays that could not be eliminated because of only having one image per filter, as well as possible foreground dwarfs that should fall in the red supergiant region of the CMD. We also ran a second set of solutions, adopting a cap of 1 for the maximum allowed value of $M_i - N_i \ln M_i$, which is perhaps an overly drastic but very effective way of eliminating the contribution from bad points to our fitting function.

In order to estimate the effects of distance and extinction uncertainties, the simulations were run with a range of allowed values. We found that models with slightly higher extinction (and a correspondingly lower distance modulus) than our assumed value of $A_V = 0.11$ from the Schlegel et al. (1998) maps were preferred, possibly as a result of internal reddening within UGC 4483.  Our preferred distance modulus from the CMD fitting is $\mu_0 = 24.45 \pm 0.11$, with an extinction of $A_V = 0.16 \pm 0.05$, both values consistent with the Schlegel et al. (1998) extinction and RGB tip distance modulus calculated in Section 3.1. Because of the large uncertainties involved in the CMD-fitting technique, we choose to not revise our previous distance, but instead note the possibility that internal reddening within UGC 4483 would result in a distance closer by up to 0.1 magnitudes.

Finally, we estimate the star formation history of UGC 4483 as a function of metallicity and age by adopting all good fits to the observed data. The scatter between different fits will be used to estimate uncertainties due to extinction and distance, while Monte Carlo simulations were used to estimate uncertainties due to small number statistics in the Hess diagrams. Our resulting star formation history is shown in Table \ref{tabSFH}.
\begin{table}
\begin{center}
\caption{Star formation history and age-metallicity relation of UGC 4483.}
\label{tabSFH}
\begin{tabular}{cccccc}
Start & End & SFR & $\logz$ & $\logz$ \\
Gyr & Gyr & $10^{-3}M_{\odot}$yr$^{-1}$ & mean & spread \\
\hline
 0.1 & 0.0 & $3.3 \pm 2.2$ & $-1.7 \pm 0.5$ & $0.4 \pm 0.2$ \\
 0.3 & 0.1 & $1.4 \pm 0.5$ & $-0.9 \pm 0.4$ & $0.2 \pm 0.1$ \\
 2.0 & 0.3 & $0.0 \pm 0.5$ & $-1.3 \pm 0.4$ & $0.3 \pm 0.2$ \\
15.0 & 2.0 & $1.3 \pm 0.3$ & $-1.4 \pm 0.4$ & $0.4 \pm 0.1$ \\
\end{tabular}
\end{center}
\end{table}

As anticipated, it was difficult to determine the star formation rate between 0.3 and 2.0 Gyr ago, due to the small number of such stars falling above the limits of our photometry. Otherwise, we were able to measure a statistically significant amount of star formation during each epoch (which was expected, of course, due to the presence of the corresponding CMD features). We find a time-averaged mean star formation rate of $1.3 \pm 0.2 \times 10^{-3} M_{\odot}$yr$^{-1}$ for UGC 4483, with a recent star formation rate (in the past 100 Myr) perhaps a factor of 2--3 higher than that value.

We find the mean metallicity of UGC 4483 to be $\logz = -1.3 \pm 0.2$, consistent with the value of $\logz = -1.4$ estimated from the spectroscopy of Skillman (1991). There is no evidence of significant chemical enrichment, with the mean metallicity in the past 300 Myr $-1.2 \pm 0.3$ and that from 2 to 15 Gyr ago to be $-1.4 \pm 0.3$. This is consistent with the observation in Section 4.1 that the single-metallicity isochrones (at $\logz = -1.4$) appeared to fit all phases of stellar evolution quite well.  However, we do find a metallicity spread of a few tenths of a dex throughout the galaxy's history, consistent with the observation of the broad RGB and the ``metal-rich'' AGB in the CMD.

\section{Discussion}

\subsection{Comparison with Similar Objects}
In addition to UGC 4483, WFPC2 observations have also been made of two similar nearby ($< 5$ Mpc) galaxies: UGC 6456 (VII Zw 403, Lynds et al. 1998) and UGC 8091 (GR8, Dohm-Palmer et al. 1998). We give a comparison between their properties in Table \ref{tabComp}.
\begin{table}
\begin{center}
\caption{Comparison of three blue compact galaxies imaged by WFPC2.}
\label{tabComp}
\begin{tabular}{cccc}
Parameter & UGC 4483 & UGC 6456 & UGC 8091 \\
\hline
$\mu_0$            & 27.53          & 28.25$^1$      & 26.65$^2$ \\
Distance           & 3.2 Mpc        & 4.5 Mpc$^1$    & 2.1 Mpc$^2$ \\
Integrated $B$     & 14.95$^3$      & 14.28$^4$      & 14.46$^4$ \\
Integrated $(B-V)$ &  0.44$^3$      &  0.38$^4$      &  0.33$^4$ \\
Extinction ${A_B}^5$ &  0.15        &  0.16          &  0.11 \\
Integrated $M_B$   & $-12.73$       & $-14.13$       & $-12.30$ \\
Dimensions (arcmin)$^6$ & 1.4$\times$0.7 & 1.6$\times$0.9 & 1.1$\times$0.9 \\
Dimensions (kpc)   & 1.3$\times$0.7 & 2.1$\times$1.2 & 0.7$\times$0.6 \\
Youngest Stars     & 10 Myr         & 4 Myr$^1$      & $<$10 Myr$^2$ \\
$\logz$ of Stars   & $-1.3$         & $-1.3^1$       & $-1.3^2$ \\
\hline
\end{tabular}
\end{center}
$^1$Lynds et al. (1998) \\
$^2$Dohm-Palmer et al. (1998); distance calculated from their measurement of the RGB tip \\
$^3$Makarova (1999) \\
$^4$Hopp \& Schulte-Ladbeck (1995) \\
$^5$Schlegel et al. (1998) \\
$^6$From the UGC catalog \\
\end{table}

The three objects are quite similar in their properties, all with colours between $(B-V)$ of 0.33 and 0.44, current or very recent star formation, and current metallicities of roughly 5\% solar ($\logz \simeq -1.3$). As noted in Section 4.1, the stellar populations of UGC 4483 and UGC 6456 have similar features, most notably the relatively metal-rich AGB and the broad red giant branch, implying a presence of at least a small number of relatively metal-rich ($\logz \simeq -0.9$) stars in both objects. It is also of interest that none of the objects were detected by the 2MASS survey or by the IRAS 60$\mu$m survey.

\subsection{Summary}
We have presented WFPC2 F606W and F814W observations of UGC 4483, an irregular galaxy in the M81 / NGC 2403 complex. Stellar photometry was made using HSTphot, showing a clear upper main sequence, red and blue supergiants, the red giant branch tip, and an asymptotic giant branch. We measure a true distance modulus of $\mu_0 = 27.53 \pm 0.12$ (corresponding to a distance of $3.2 \pm 0.2$ Mpc) from the RGB tip, although there is possibly a non-negligible amount of internal absorption within UGC 4483 that would cause the distance to become closer by a tenth of a magnitude. This distance places UGC 4483 as a member of the NGC 2403 subgroup, and indicates that the galaxy has similar integrated properties to those of UGC 6456 (VII Zw 403) and UGC 8091 (GR8), which have also been observed by WFPC2.

We have been able to resolve some of the brightest stars of a young cluster near the northern edge of the galaxy, and find an age of $\sim 10-15$ Myr and initial mass of $\sim 10^4$ $M_{\odot}$. Its brightest resolved star is a blue supergiant, with the remaining resolved stars all falling on the main sequence. We measure integrated magnitudes of $V = 18.66 \pm 0.21$ and $I = 18.54 \pm 0.10$ for the stellar contribution, with the large uncertainty in $V$ dominated by uncertainties in the subtraction of line emission through the F606W filter.

We find spatial differences in the distribution of main sequence stars, blue loop stars, red supergiants, and RGB/AGB stars. The main sequence stars are the most tightly concentrated, and are mostly found in the half of the galaxy containing the cluster. Blue loop stars are found throughout the main body of the galaxy, while RGB/AGB stars show the largest dispersion. We cannot discern whether this trend in distribution results from a recent localization of star formation or from increased migration of older stars.

Finally, we made a rough estimate of the galaxy's star formation history through a numerical fit to the colour-magnitude diagram. Accurate measurement of the star formation rates from 0.3 to 2 Gyr ago is difficult, due to the small number of such stars that would fall above our photometry limit. However, we found statistically significant detections of stars in the other age ranges, measuring a mean metallicity of $\logz = -1.3 \pm 0.2$ and time-averaged star formation rate of $1.3 \pm 0.2 \times 10^{-3} M_{\odot}$yr$^{-1}$.

\section*{Acknowledgments}
We appreciate the help of Leo Girardi in preparing a complete set of isochrones for the population synthesis work and for providing new isochrones for $\logz = -2.3$. AED acknowledges support for this work from NASA through grant numbers GO--07496 and GO--02227.06--A from the Space Telescope Science Institute, which is operated by AURA, Inc., under NASA contract NAS 5--26555. DG acknowledges financial support for this project received from CONICYT through Fondecyt grants 1000319 and 8000002, and by the Universidad de Concepci\'on through research grant No. 99.011.025.

\label{lastpage}
\end{document}